\documentclass[prb,twocolumn,showpacs,floatfix]{revtex4}
\usepackage{graphicx}
\usepackage{dcolumn}
\usepackage{amsmath}

\begin{document}
\title{Wigner molecules in quantum dots: a quantum Monte Carlo study}

\author{A. Harju} \author{S. Siljam\"aki} \author{R.M. Nieminen}

\affiliation{Laboratory of Physics, Helsinki University of Technology,
P.O. Box 1100, 02015 HUT, Finland}

\date{\today}

\begin{abstract}
We study two-dimensional quantum dots using the variational quantum
Monte Carlo technique in the weak-confinement limit where the system
approaches the Wigner molecule, i.e., the classical solution of point
charges in an external potential. We observe the spin-polarization of
electrons followed by a smooth transition to a Wigner-molecule-like
state as the confining potential is made weaker.
\end{abstract}

\pacs{73.21.La, 71.10.-w}

\maketitle

\section{Introduction}
Semiconductor quantum dots (QD) are small devices containing a tunable
number of electrons in an external confinement
potential.\cite{leo_rev} The significant progress in the fabrication
of these devices during the last few years\cite{leo_s} has stimulated
an increasing interest in investigating the properties of such
systems.  From the theoretical point of view, QDs are ideal
many-electron objects for the study of fundamental physical properties
of correlated electrons.

Perhaps the most striking feature in these artificial atoms is that
the system parameters can easily be changed, unlike in real atoms
where the parameters are natural constants. Also the typical length
scales of interactions and confinement are equal, which should in
principle make the correlation effects more enhanced compared to
normal atoms. However, the experimentally observed states of QDs in
weak magnetic fields can easily be understood as single configurations
of non-interacting one-particle states.\cite{leo_s,wB} One might
suppose that the lack of a strong central Coulomb potential (which
organizes the states of normal atoms to shells) would make it more
difficult to identify the electronic structure of artificial atoms. It
turns out, however, that the important role of the central potential
is taken by the symmetries of the many-body wave function.\cite{wB,sB}
Consequently the understanding of the topology of the many-body wave
function is of central importance.

A most suitable method for studying the many-body wave functions is
the variational quantum Monte Carlo (VMC) technique. It is intimately
coupled to the structure of the wave function, and the use of the
variational principle enables one to find the best form. The exact
diagonalization (ED) method is useful for the studies of small
electron numbers\cite{wB} or for finding eigenstates in some important
subspace of the full Hilbert space, as in the studies of electrons in
such a strong magnetic field that the physics is mainly determined by
the single-particle states of the lowest Landau level.\cite{sB} ED has
been used for QDs by many authors, see Ref.~\onlinecite{Jdiag} and
references therein.  Mean-field methods such as the spin-density
functional theory (SDFT) are especially useful when the number of
particles is large.\cite{Jdiag}

In this work, we first compare our VMC results with the most accurate
energies in the literature, showing that a rather intuitive wave
function results in extremely accurate energies. Then we concentrate
on the weak confinement limit of a six-electron QD.  One should note
that making confinement weaker makes the density smaller.  In this
limit the QD approaches the classical solution of point charges in the
QD potential.  The most accurate results for this limit are restricted
to rather strong confinement\cite{Jdiag}, where no trace of Wigner
molecule or spin-polarization was found.  For weaker confinements than
considered in Ref.~\onlinecite{Jdiag} we find that the system spin-polarizes
before the electrons localize around the classical positions.  We also
introduce a conditional probability density which shows to be a good
measure of the extent of the localization of the electrons.  In
addition, we find an interesting independent-electron-type scaling of
the energy for a large range of confinement strengths.

\section{Model and Method}
The commonly used Hamiltonian for $N$ electrons in a two-dimensional
QD can be written as
\begin{equation}
H = \sum _{i=1}^N\left ( -\frac{\hbar^{2}}{2 m^{*}} \nabla_{i}^{2}
+ \frac {m^{*}\omega^{2}}{2}r_{i}^{2}
\right ) + \sum_{i < j} \frac {e^{2}}{\epsilon
r_{ij}} ,
\label{ham}
\end{equation}
where 
$\omega$ is the strength of the external confining potential, and
the effective mass of the electrons $m^*$ and the dielectric constant
$\epsilon$ are used to model the properties of the semiconductor
material studied.  For GaAs, the material parameters are $m^*/m_e =
0.067$ and $\epsilon=12.4$.\cite{units} We assume that the electrons
move in the $z=0$ plane and omit the magnetic field in this study.

In this work, we use variational wave functions of the form
\begin{equation}
\Psi = 
D_{\uparrow} D_{\downarrow}
\prod_{i<j}^N J(r_{ij}) \ ,
\label{wf}
\end{equation}
where the two first factors are Slater determinants for the two spin
types, and $J$ is a Jastrow two-body correlation factor. We neglect
the three-body and higher correlations. For the Jastrow factor we use
\begin{equation}
J(r)=e^{\frac{C r}{a+b r}} \ ,
\end{equation}
where $a$ is fixed by the cusp condition to be 3 a for pair of equal
spins and 1 for opposite ones and $b$ is a parameter, different for
both spin-pair possibilities.  We take the wave function to be real,
as we have neglected the magnetic field. The single-particle states
$\psi$ are expanded in the basis of gaussians as
\begin{equation}
\psi({\bf r}) =\sum_{i=1}^{N_s} c_i H_{n_{x,i}}(\hat x) H_{n_{y,i}} (\hat y) e^{- \frac 12 \hat r^2} \ ,
\end{equation}
where $H_n$ is a Hermite polynomial of order $n$ and 
$\hat {\bf r}=(\hat x, \hat y)=(c_{x} x,c_{y} y)$, where
 $c_x$ and $c_y$ scale
coordinates.

We calculate the energy $E$ which is bound by the exact energy $E_0$
using
\begin{equation}
E_0 \le E = \lim_{M \to \infty} \frac {1}{M} \sum_{i=1}^{M} E_{\rm L} ({\bf
R}_i) \ ,
\end{equation}
where the local energy is $E_{\rm L} = H \Psi / \Psi$, and depends on
the electron configuration ${\bf R}$. The configurations ${\bf R}$ are
distributed according to $| \Psi |^2$.  We optimize the variational
parameters using the stochastic gradient approximation
(SGA).\cite{sga} The SGA optimization method involves stochastic
simulation in two spaces: the configuration and the parameter
space. In the configuration space, a set of $m$ configurations $\{
{\bf R}_j \}$ is sampled from a distribution $|\Psi(\mbox{\boldmath
$\alpha$})|^2$, where $\mbox{\boldmath $\alpha$}$ is the current
parameter vector.  In the parameter space, the parameters at iteration
$i+1$ are obtained from the previous ones by the formula
\begin{equation}
\mbox{\boldmath $\alpha$}_{i+1}=\mbox{\boldmath $\alpha$}_i-\gamma_i
\nabla_{\mbox{\boldmath $\alpha$}} {\cal Q}_i \ ,
\label{SGA_formula}
\end{equation}
where $\gamma_i$ is a scaling factor of the step length and ${\cal Q}$
is an approximation to the cost function. For energy minimization the
cost function is simply the mean of the local energies over the set
of configurations:
\begin{equation}
{\cal Q}=\langle E_{\rm L} \rangle= \frac 1m \sum_{j=1}^m E_{\rm
L}({\bf R}_j) \ .
\end{equation}
The scaling factor $\gamma$ has an important role in averaging out the
fluctuations in the approximate gradient, ensuring the convergence. On
the other hand, too small a value of $\gamma$ would overdamp the
simulation.  If one uses a sequence $\gamma_i\propto i^{-\beta}$, one
should have $\frac 12 < \beta \le 1$.

Lin {\it et al.}\cite{anad} have shown that in the case of real wave
functions and energy minimization, the derivative of the energy $E$
with respect to a variational parameter $\alpha_i$ is simply
\begin{equation}
\frac{\partial E}{\partial \alpha_i}=2 \left\{ \left\langle E_{\rm L} \times
\frac{\partial \ln \Psi}{\partial \alpha_i} \right\rangle -E \times
\left\langle \frac{\partial \ln \Psi}{\partial \alpha_i} \right\rangle
\right\} \ ,
\label{anaa}
\end{equation}
where the average $\langle \dots \rangle$ is over the whole Metropolis
simulation.\cite{anad} One can implement this simple formula also for
the SGA algorithm, with the small modification that the average is
taken over only the current set of $m$ configurations.

\section{Results}
\subsection{Comparison with other approaches}
Before presenting the results for the Wigner-molecule-limit of a
six-electron QD, we compare our VMC results with the most accurate QD
results up-to-date.  In a recent VMC and diffusion quantum Monte Carlo
(DMC) study,\cite{dmc} Pederiva {\it et al.} study quantum dots using
a similar model as we do. They find the accuracy of VMC to be rather
good compared to DMC, except in the case of three electrons.  For this
case, we have first taken the single-particle states to be the
non-interacting ones with quantum numbers $(0,0)$ and $(1,0)$.  With
the GaAs choice of system parameters given above, the confinement
energy corresponding to the study of Pederiva {\it et al.}\cite{dmc}
is around $3.32$~meV.  The total energy is found to be
$26.563(1)$~meV\cite{uns} which is reasonably close to the DMC value
of Pederiva {\it et al.}, namely $26.488(3)$~meV.\cite{dmc} On the
other hand, the VMC energy reported by Pederiva {\it et al.} is
$29.669(3)$~meV.\cite{dmc}  Optimizing the exponentials lowers our
VMC energy to $26.5406(8)$~meV.  The difference of our VMC energy to
the DMC one is around 0.05~meV which is small compared to the SDFT
error $\sim$ 0.4~meV.\cite{dmc}  For the {\em six}-electron case, the
energy with non-interacting single-particle states is found to be
$90.27(1)$~meV, which is again closer to the DMC energy 90.11(1)~meV
of Pederiva {\it et al.}\cite{dmc} than the VMC one 90.368(4)~meV.
By freeing the parameters in the single particle states one does not
lower the energy within the statistical error, and the optimal values
are thus equal to the ones with the non-interacting states. This is a
very important result, showing that the change in the wave function
introduced by the Coulomb interaction is very accurately taken into
account by the two-body Jastrow factors used.  The reason why the
optimization of the single-particle states was important in the
three-electron case is most probably that there the number of spin-up
and spin-down electrons is different.

It is also interesting to compare the results obtained with VMC with
those of Reimann {\it et al.} for the six-electron case.\cite{Jdiag}
In their study, they use both SDFT and ED and consider six electrons
with various strengths of the confining potential. We compare four
different strengths of confinement, corresponding in their work to the
cases of $r_s=$1, 2, 3, and 4 $a_B^*$.  Their ED energies are given in
Table~\ref{Jt}
\begin{table}
\caption{Total energy (in meV) of the six-electron dot for different
confinements and the two spin states. 
\label{Jt}}
\begin{ruledtabular}
\begin{tabular}{ccccc}
 & \multicolumn{2}{c}{S=0} & \multicolumn{2}{c}{S=3}\\
$\hbar \omega$ & VMC & ED\footnotemark[1] & VMC & ED\footnotemark[1] \\ \hline
7.576 & 168.90(1) & 169.2& 180.40(1) & 180.5\\
2.678 &  76.91(1) & 77.17& 79.271(4) & 79.38\\
1.458 & 49.101(5) & 49.35& 49.934(3) & 50.10\\
0.947 & 35.864(3) & 36.11& 36.231(2) & 36.41
\end{tabular}
\end{ruledtabular}
\footnotetext[1]{From Ref.~\onlinecite{Jdiag}.}
\end{table}
with our VMC energies. One can see that the difference in the energies
is between 0.1 and 0.3~meV, the VMC energies being lower. This
comparison shows that the finite basis used by Reimann {\it et al.} is
too restricted to describe the many-body wave function accurately.
This comparison also shows that the results obtained with VMC are very
accurate for the particle number in question.

We also compare the accuracy of our VMC results with the path-integral
Monte Carlo simulations of Reusch {\it et al.} for
$N=8$.\cite{pimc} As we are below mainly interested in the fully
spin-polarized states, we compare only the $S=4$ energies.  This is
not a closed-shell case, and our variational wave function for the two
highest states is constructed from the four states with $n_x + n_y=3$.
For interaction strength $C=2$, their energy is 48.3(2)~$\hbar \omega$
which is slightly higher (but within error bars) than our energy of
48.201(1)~$\hbar \omega$. As one can see, the statistical error is two
orders of magnitude smaller in our result.  For their most strongly
interacting case, namely $C=8$, Reusch {\it et al.} obtain energy
103.26(5)~$\hbar \omega$ which is again less accurate than our energy
103.137(1)~$\hbar \omega$.  Also this test shows that our results are
very accurate in the limit of strong interaction.

One can conclude from these comparisons that the wave function used is
very accurate. Most striking is the observation that after the Jastrow
factor is added to the wave function, the remaining problem can be
treated using non-interacting single-particle states. Usually in VMC
studies this remaining problem contains some interactions which are
taken into account in a mean-field fashion by modified single-particle
states.

\subsection{Classical limit}

Next, we study the transition to the classical limit in the
six-electron case. The ground-state structure of six purely classical
point charges in a parabolic potential minimizes the energy
\begin{equation}
E_{CL}=\frac 12 \sum_i r_i^2 + \sum_{i<j} \frac{C}{r_{ij}} \ ,
\end{equation}
where we have used reduced units\cite{units}. If we keep other
parameters fixed and change only the confinement strength $\omega$ (as
we do below), one can see that the interaction strength $C$ scales as
$C \propto \omega ^ {-1/2}$.  The minimum-energy positions of
electrons form a pentagon around one electron at the center. One can
find the scaling of the classical cluster size and energy by writing
the coordinates as ${\bf r}=r_c {\bf \hat r}$, where coordinates $\hat
r$ are fixed and the scaling is in $r_c$.  This results for the energy
\begin{eqnarray}
E_{CL}&=&r_c^2 \frac 12 \sum_i \hat r_i^2 + \frac{C}{r_c} \sum_{i<j}
\frac{1}{\hat r_{ij}} \nonumber\\
&=& r_c^2 V_1 + \frac{C}{r_c} V_2\ ,
\end{eqnarray}
where $V_1$ and $V_2$ are constants, and solving ${\partial
E_{CL}}/{\partial r_c}=0$ results in $r_c=(C V_2/V_1)^{1/3}\propto
\omega^{-1/6}$. Thus the energy scales (in units of $\omega$) as
$E_{CL}\propto \omega^{-1/3}$. The minimum energy $E_{CL}^*$ can be
found to be $E_{CL}^*[{\rm meV}] \approx 30.46(\hbar\omega[{\rm
meV}])^{2/3}$.

\begin{figure}[htb]
\includegraphics[width=8cm]{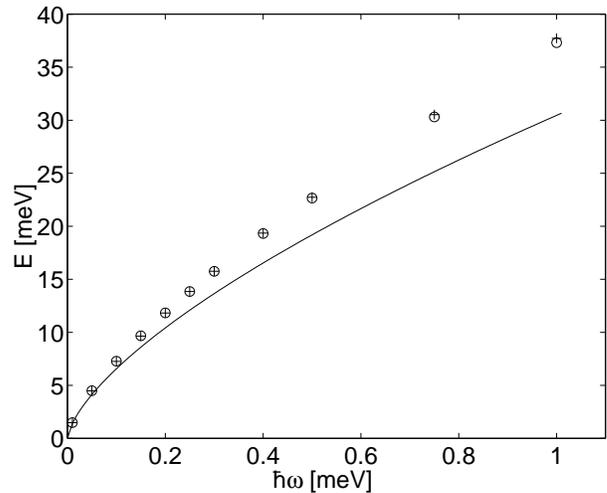}
\caption{Total energy for spin states $S=3$ (marked with pluses) and
$S=0$ (circles) as a function of $\hbar \omega$. The line presents the
classical energy $E_{CL}^*$.  }
\label{Eee}
\end{figure}

\begin{figure}[htb]
\includegraphics[width=8cm]{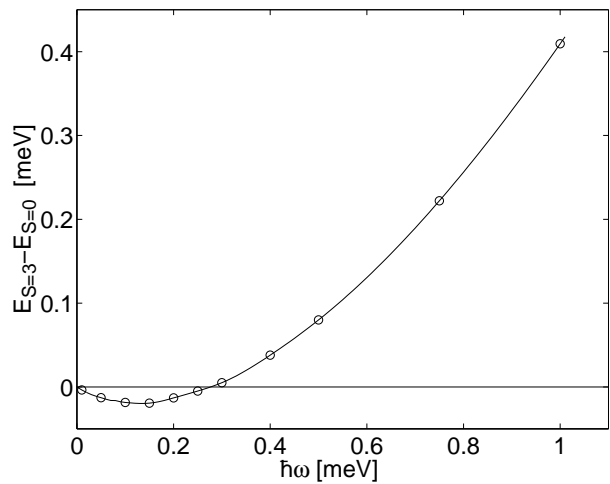}
\caption{Energy difference between the spin states $S=3$ and $S=0$ as
a function of $\hbar \omega$. The line is to guide the eye.}
\label{xtraa}
\end{figure}

\begin{figure}[hbt]
\includegraphics[width=8cm]{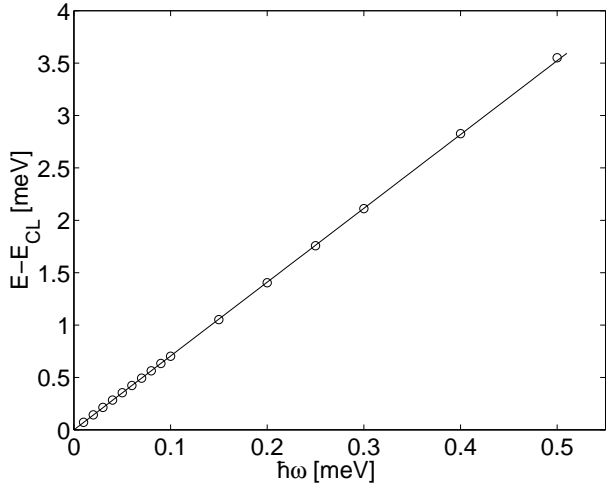}
\caption{Remaining energy for $S=3$ when $E_{CL}^*$ (see text) is
subtracted. The line presents a linear fit to points $\hbar \omega \le
0.1$~meV.}
\label{xtrab}
\end{figure}

The quantum-mechanical energies resulting from our VMC calculations
for two spin polarizations are presented in Fig.~\ref{Eee}
with the classical energy $E_{CL}^*$.
One can see that the quantum-mechanical energies are very close to
each other, especially in the small $\hbar \omega$ limit, where the
energies also approach the classical one.  The difference
between the quantum-mechanical energies can be seen more clearly in
Fig.~\ref{xtraa},
where we have plotted the energy difference between the fully- and
non-polarized states.  One can see that spin-polarization is predicted
at $\hbar \omega \approx 0.28$~meV.  We have not found ground states
with partial spin-polarization.  Below, we concentrate on the fully
spin-polarized case.  If one subtracts from the total energy the
minimum value of the classical potential energy $E_{CL}^*$, the
remaining energy has a linear behavior at small $\hbar \omega$ as
shown in Fig.~\ref{xtrab}.
One way of understanding this linear behavior is to consider first the
limit $\hbar \omega \to 0$. In this limit, the electrons localize to
the classical positions. If $\hbar \omega$ is now made larger, the
electrons start to oscillate around the classical positions.  A first
approximation for this is to assume that each electron is in a
harmonic potential $\tilde V({\bf r}_i) = \frac 12 \tilde \omega^2
r_i^2$, with the strength scaling as $\tilde \omega \propto \omega$ as
a function of $\omega$. As a result of this, the total energy has,
apart from the classical energy, a zero-point energy that also scales
as $\tilde \omega$. Thus we obtain a linear behavior for the
difference between the total energy and the classical energy similarly
as in Fig.~\ref{xtrab}.  The error in the linear fit of
Fig.~\ref{xtrab} is smaller than 3~$\mu$eV for $\hbar \omega \le
0.1$~meV where the fit is made, and for $\hbar \omega=1$~meV still
only 0.2~meV, which is of the same order as the difference in energy
between the VMC and ED results in Table~\ref{Jt}. For $\hbar
\omega=2$~meV the error is around 1~meV.  Thus the approximation
derived above works surprisingly well even for a rather strong
confinement. One should note that the approximation did not contain
any information of the interaction between electrons (apart from the
classical potential energy), and thus the system can be seen
energetically as a collection of nearly independent electrons
oscillating in an effective potential.

The most probable configuration ${\bf R}^*$, maximizing the density $|
\Psi({\bf R}) |^2$ should approach in the limit of weak confinement
the classical electron positions. This is not, however, enough to show
that the system is close to a classical one. One can study the quantum
fluctuations very conveniently using the conditional single-particle
probability distribution $\tilde \rho ({\bf r})$, defined as
\begin{equation}
\tilde \rho ({\bf r}) = 
\left | 
 \frac{\Psi({\bf r},{\bf r}^*_2,\dots,{\bf r}^*_{N})}{\Psi({\bf r}_{1}^{*},{\bf r}^*_2,\dots,{\bf r}^*_{N})}
\right |^2 \ ,
\label{rhod}
\end{equation}
where the coordinates ${\bf r}_i^*$ are fixed to the ones from the
most probable configuration ${\bf R}^*$. In the classical limit, the
density $\tilde \rho ({\bf r})$ is more and more peaked around the
classical position ${\bf r}_1^*$, but still shows quantum
fluctuations.  For the two-electron case, this is very similar to the
conditional probability distribution used for a two-electron
QD.\cite{2ehf} One difference is that in $\tilde \rho$ the fixed
electron is at the most probable position, which is in our opinion the
most natural choice.  The most important advantage of $\tilde \rho$ is
that it shows much more clearly the amount of localization for larger
particle numbers than the conditional probability distribution. The
reason for this is that one usually fixes only one electron in
constructing the conditional probability distribution, and when the
particle number is large, the effect of one fixed electron gets
smaller, and the rest of electrons can, for example, show collective
motion that conceals the localization. If, on the other hand, one
fixes all but one electron, the most natural choice is $\tilde
\rho$. The calculation of $\tilde \rho$ is very easy, especially in
VMC.  One should first, of course, find the most probable electron
positions.  In doing this, the gradient of the wave function (needed
also for the calculation of the local energy, and for this reason
usually done analytically) is very useful. After that, one moves the
``probe electron'' to all points where the value of $\tilde \rho$ is
wanted, and evaluates the ratio of wave functions as in
Eq.~(\ref{rhod}).  This ratio is automatically done when sampling the
configurations in a VMC simulation.  One should also notice that
$\tilde \rho$ does not contain noise unlike many more common VMC
observables, such as the density or the radial pair distribution
function.

\begin{figure*}[hbt]
\includegraphics[width=6cm]{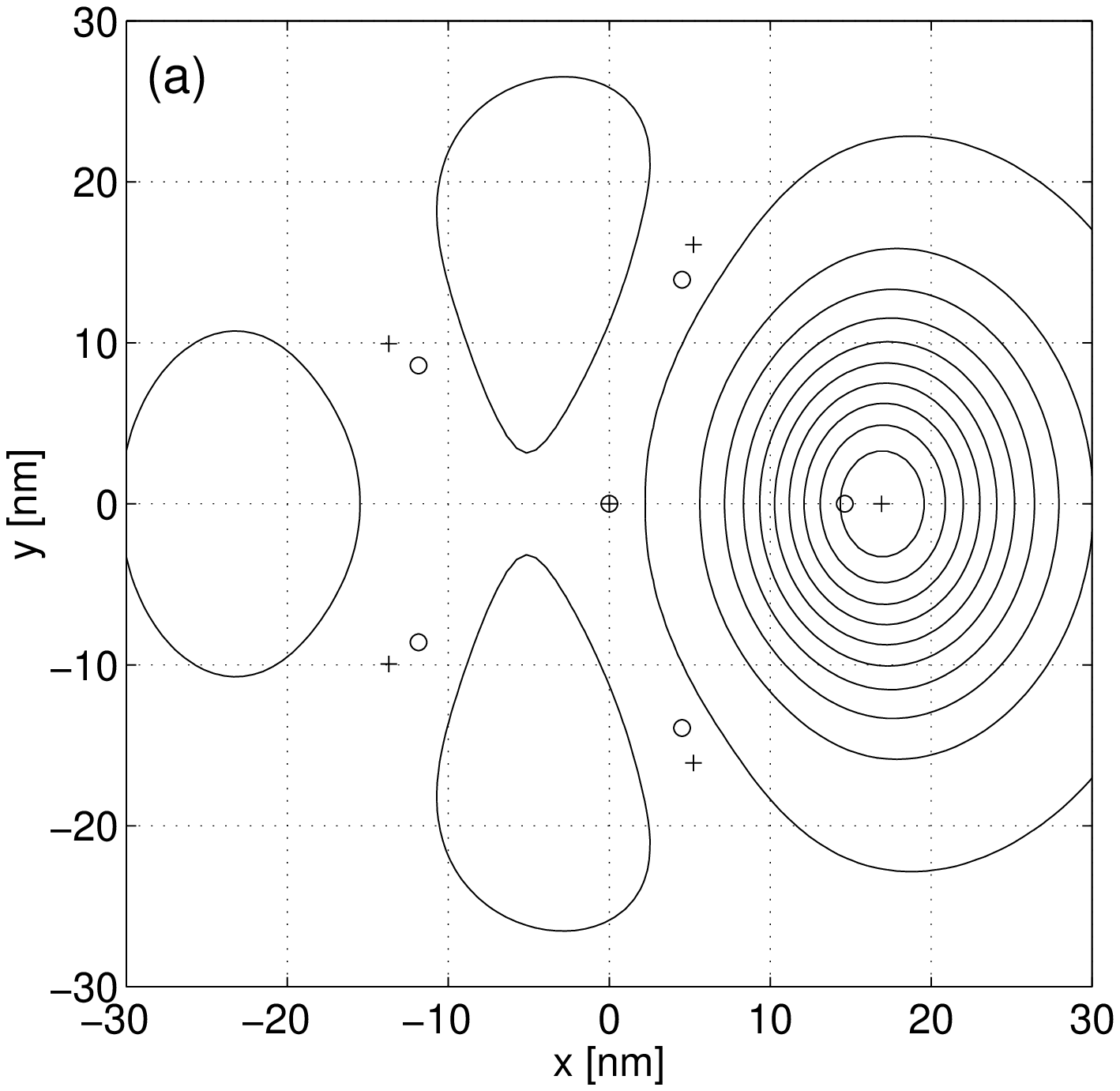}
\includegraphics[width=6cm]{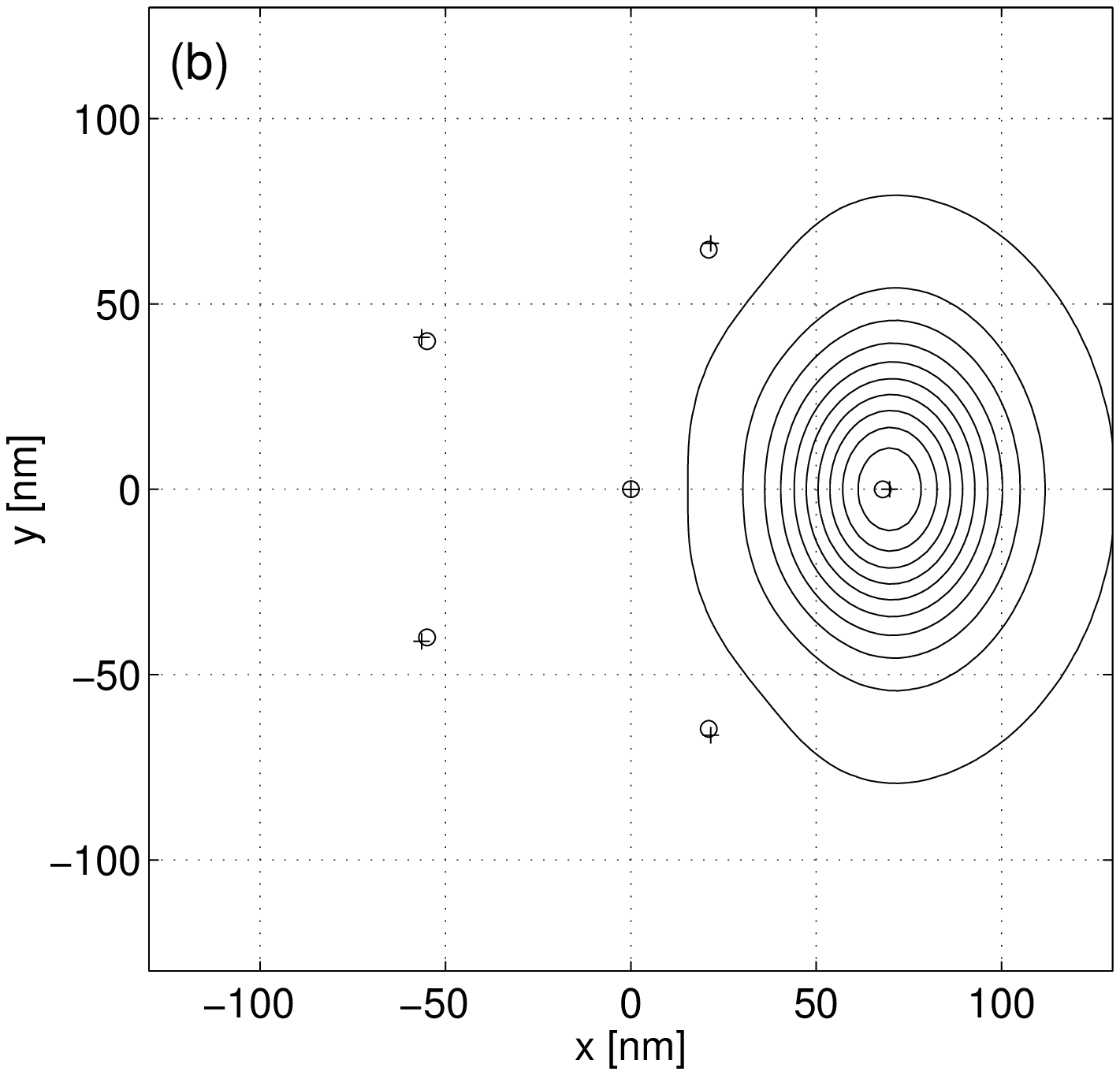}\\
\includegraphics[width=6cm]{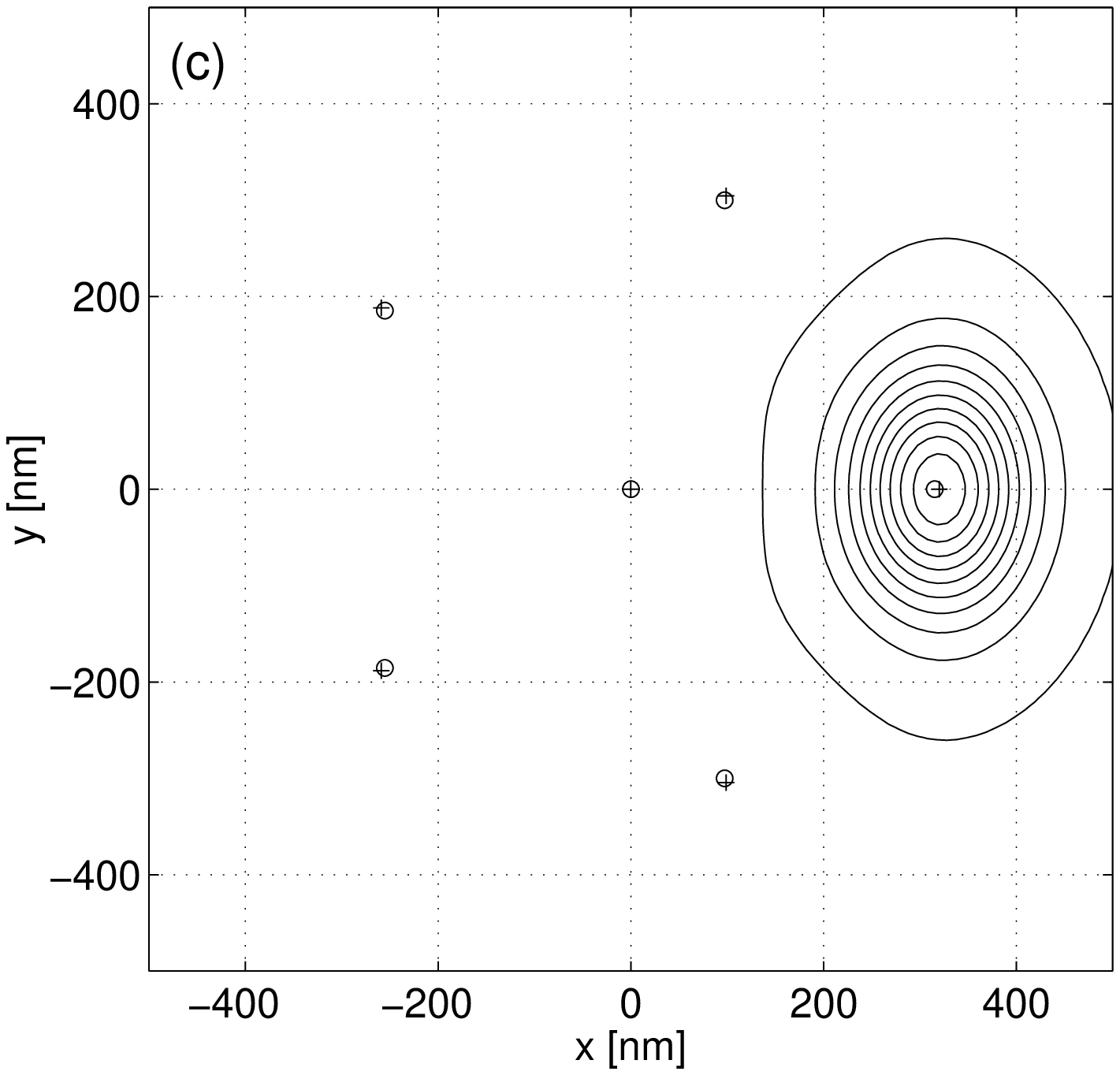}
\includegraphics[width=6cm]{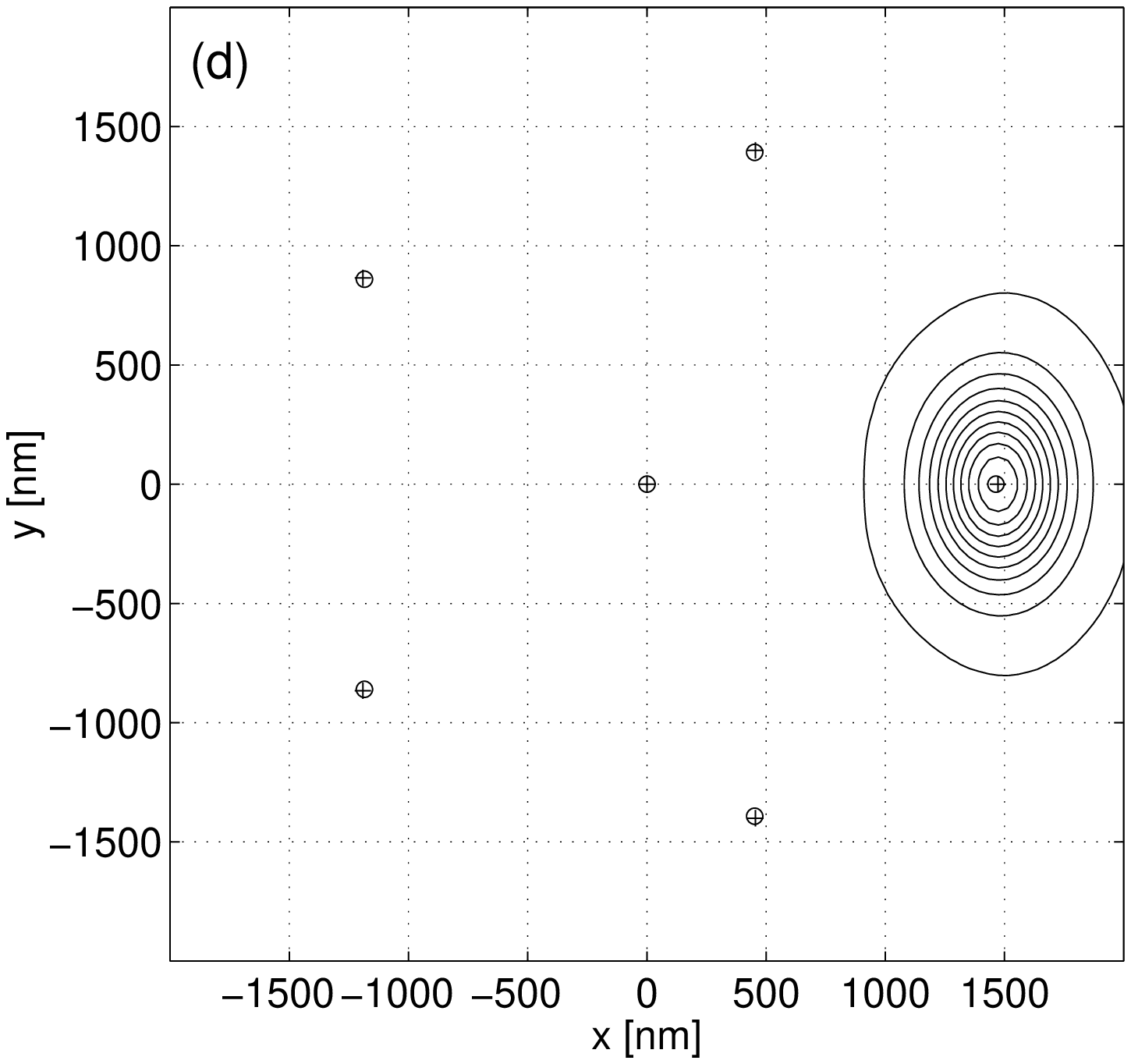}
\caption{Conditional probability density $\tilde \rho ({\bf r})$ for
the right-most electron. The contours are uniformly from 0.01 to 0.91.
We mark with a plus the most probable electron positions, and with a
circle the classical positions. The confinement strength $\hbar
\omega$ is: (a) $10$~meV, (b) $1$~meV, (c) $0.1$~meV, and (d)
$0.01$~meV.}
\label{dall}
\end{figure*}

In Fig.~\ref{dall}
we show $\tilde \rho$ for four different confinement strengths from 10
to 0.01~meV.  The most probable electron positions are for all
strengths similar to the classical ones with one electron in the
center of a pentagon. The quantum effects make the most probable
coordinates slightly larger than the classical ones, but the
difference vanishes for weak confinement.  One can also see that
$\tilde \rho$ is more localized for weaker confinement, but only the
one with $\hbar \omega=0.01$~meV looks like a Wigner molecule.  In
that case, all the points on the line $\tilde \rho=0.01$ are closer to
${\bf r}_{1}^{*}$ than any other ${\bf r}_{i}^{*}$.  Also the snapshot
animations of the systems during the simulations\cite{animation} show
the case of $\hbar \omega=0.01$~meV to resemble what one expects for a
Wigner molecule.  As we work with a finite system, we do not have a
real phase transition between $\hbar \omega=0.1$ and 0.01~meV.  In a
two-dimensional electron gas, the phase transition to a Wigner crystal
has been suggested to happen at the density $r_s \approx 37 \pm
5$\cite{2d} (in units of effective Bohr radius $a_B^*$, which is
around 9.79 nm for our system parameters). The radius of a circle that
encloses one electron on the average is thus around 360 nm at the
transition point. This is in a good agreement with the inter-electron
distances shown in Fig.~\ref{dall}, since in Fig.~\ref{dall}(c) the
distance of edge-electrons to the center one is smaller than the
suggested critical 360 nm, and in Fig.~\ref{dall}(d) the distance is
roughly four times the critical one. The approximative relation
between $r_s$ and $\hbar \omega$ presented in Ref.~\onlinecite{Jdiag}
gives $\hbar \omega \approx 0.034$~meV for $r_s=37$, which is nicely
between the cases of Figs.~\ref{dall}(c) and (d).

\begin{figure}[hbt]
\includegraphics[width=6cm]{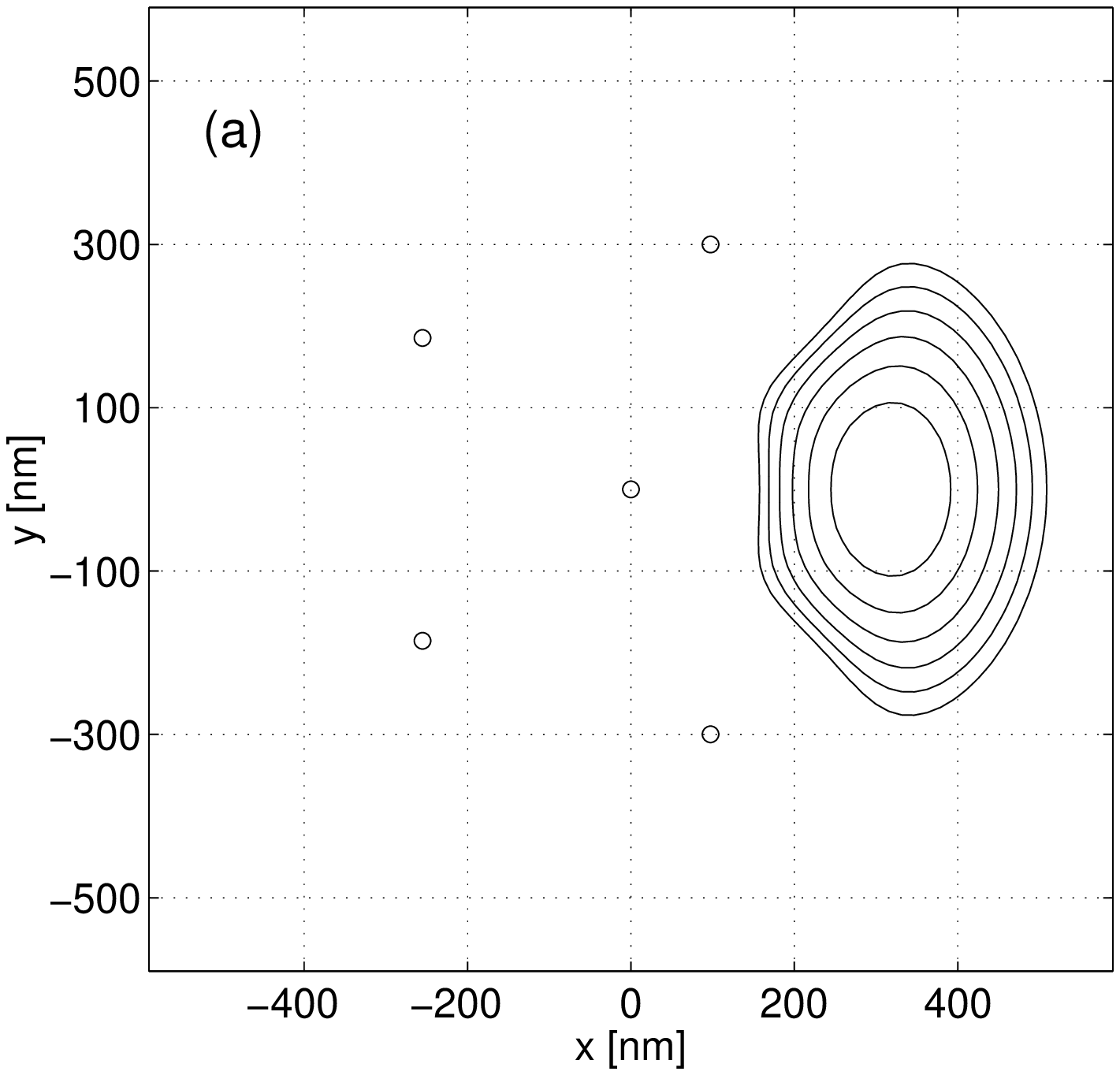}
\includegraphics[width=6cm]{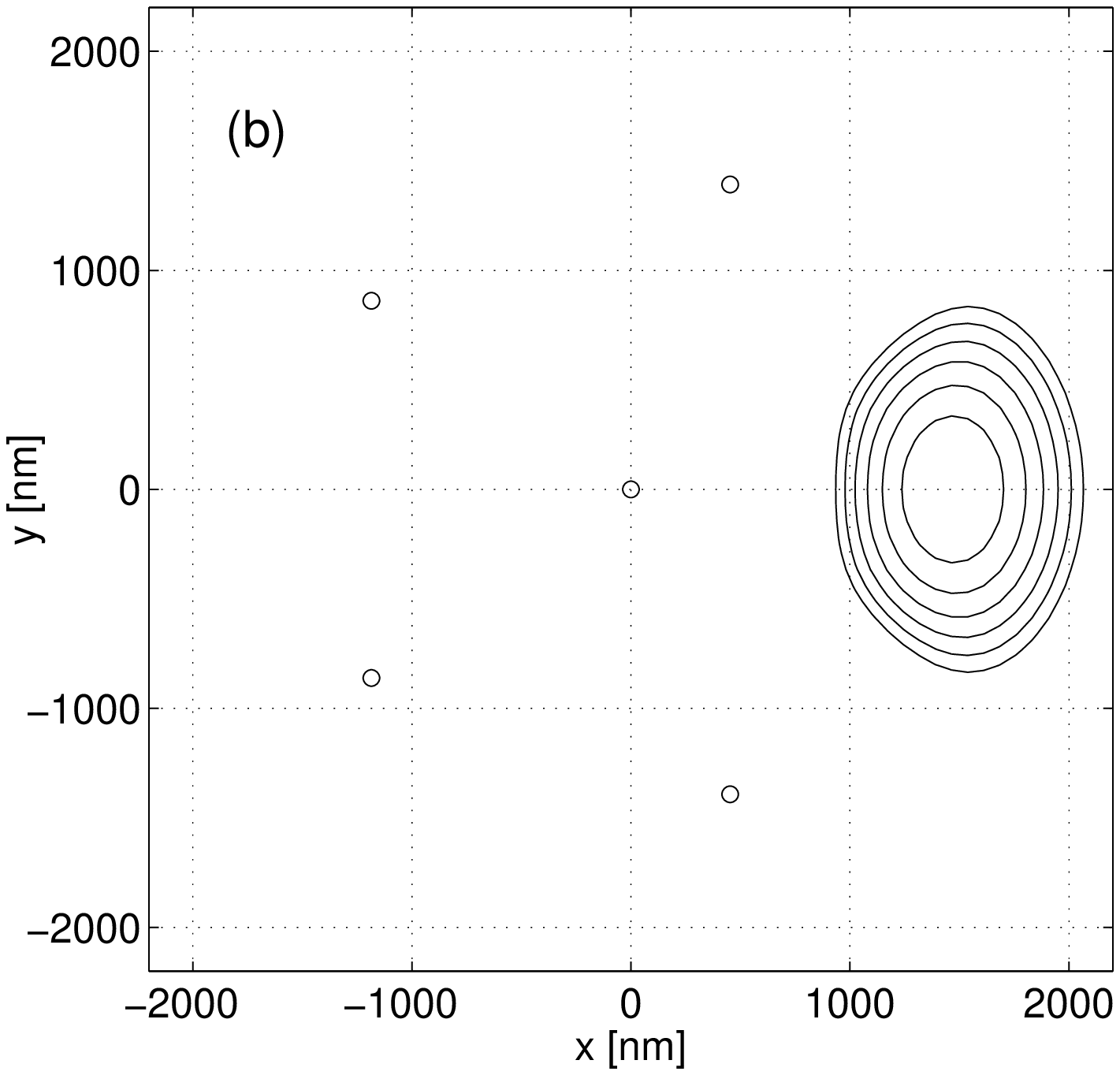}
\caption{Potentials felt by one of the electrons (contours uniformly
from half to three) when the five other electrons are on their
classical positions (marked with a circle). The confinement strength
$\hbar \omega$ is: (a) $0.1$~meV, and (b) $0.01$~meV.}
\label{pots}
\end{figure}

It is also interesting to see the
similarity between the effective single-particle potentials $\tilde V$
shown in Fig.~\ref{pots}(a) and (b),
and the corresponding $\tilde \rho$ in Fig.~\ref{dall}(c) and
(d). This similarity could be used to describe the system, as we did
above, as a system of six independent electrons, each with its own
effective potential $\tilde V$.  The assumption of parabolic $\tilde
V$ could easily be replaced by, e.g., $\tilde V = \frac 12 (\tilde
\omega_{x,i}^2 x_i^2 + \tilde \omega_{y,i}^2 y_i^2)$ keeping the
problem still solvable.  There are, however, contributions that are
not taken into account in this simple model, such as the exchange
energy which is important for the spin-polarization of the system.
Another aspect of the similarity between $\tilde \rho$ and $\tilde V$
is that it draws a connection between the wave function and the
potential in a similar fashion that is often assumed in a
semi-classical approximation.  This could be used to motivate the
studies of classical charged particles in various confinements and
also a semiclassical approach in the limit where the electrons are
close to forming a Wigner molecule.

\begin{figure}[hbt]
\includegraphics[width=8cm]{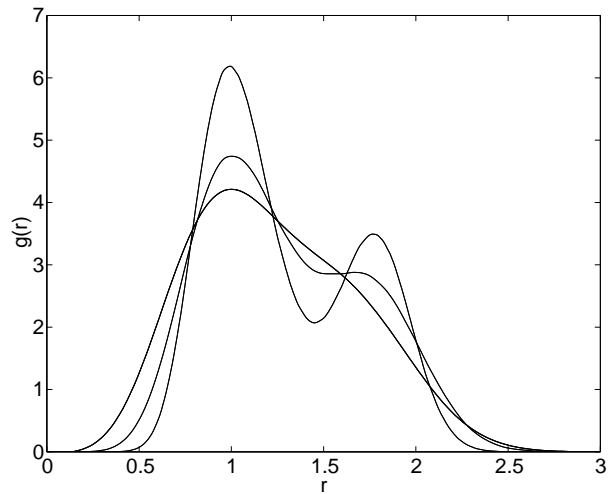}
\caption{Radial pair distribution function for $\hbar
\omega=$1.0, 0.1, 0.01~meV. The curves with smaller $\hbar \omega$ are
more peaked.  $r$ is scaled to set the first peak to one.}
\label{gee}
\end{figure}

\begin{figure}[hbt]
\includegraphics[width=7cm]{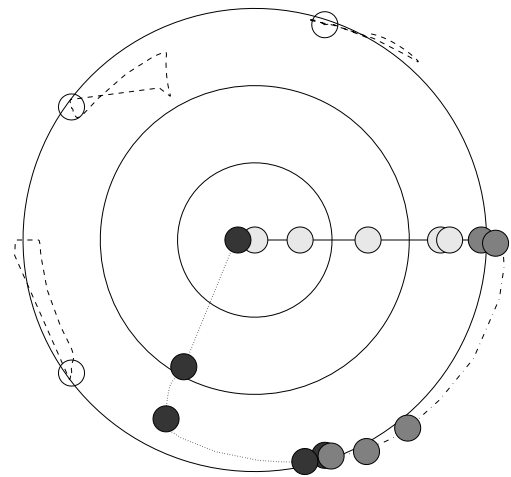}
\caption{Particle exchange for $\hbar \omega = 0.01$~meV. The electron
from the center (marked with light-gray circle) is moved to the right
along the solid line. The rest of the electrons are always at their
most probable positions. For the electrons ending at the starting
position, only these positions and the path followed is shown.  The
circles showing the distance from the center have a radius from 500 to
1500 nm.}
\label{exc}
\end{figure}

The transition to a Wigner-molecule-like regime can also be seen in
the radial pair distribution function, shown in Fig.~\ref{gee}.
The function is clearly more peaked for weaker confinement.  The peak
at $r=1$ consists of two types of electron pairs, namely ones with
both electrons on the edge, and ones with an electron in the center
and other on the edge.  The electron-electron distance in the pair of
the first type is in a classical solution 18\% longer than in the
second, the number of different pairs being the same in the classical
solution.  This double-nature cannot be seen in the first peak of
$g(r)$.  This is not surprising, as particle exchanges happen even
with $\hbar \omega=$ 0.01 meV.\cite{animation} To study exchange, we
have followed the most probable electron positions while forcing the
electron originally in the center to move to the edge. The symmetry is
broken by making small random displacements of the electrons.  The
particle exchange involves three or more electrons, as shown in
Fig.~\ref{exc}.
One can see a collective rotation of the edge electrons before one is
moved halfway to the edge. This kind of exchange is very easy for the
small QDs as the one in question.  One could argue that larger
electron numbers would make the multi-particle exchanges to a more
unlikely process.

Compared with the experimental realizations of GaAs QDs, the electron
density where the Wigner molecule is found is extremely small.  We
feel that impurities would move the transition to larger densities as
in the 2D electron gas, where the transition is found to move to
$r_s\approx 7.5 \ a_B^*$.\cite{chui} For this, the approximative
relation\cite{Jdiag} gives $\hbar \omega \approx 0.37$~meV, which is
already closer to the typical confinement strengths in experiments.
It would also be very interesting to study the effect of impurities on
the spin-polarization transition.

\section{Summary and conclusions}

We have first shown that the VMC method results energies in good
agreement with the most accurate results available.  In VMC, the
construction of the wave function clearly plays a central role.  We
have shown that the efficiency of SGA allows us to carefully optimize
also the single-particle part of the wave function, resulting VMC
energies more accurate than the previous ones where the
single-particle states are taken from mean-field approach.\cite{dmc}
On the other hand, our results show that in many cases the
non-interacting single-particle states are optimal or very close to
the optimal ones.  This is probably related to the high symmetry of
the parabolic QD and the separation of the center-of-mass and relative
motion.  We feel that the efficiency of the SGA method would be even
more useful in low-symmetry dots.

Unlike in the previous accurate study of the six-electron
QD,\cite{Jdiag} we have been able to reach low enough densities to
find a spin polarization of electrons and, in an even lower density, a
smooth transition to a Wigner-molecule-like state.  The transition
happens roughly at the same density as in the 2D
electron-gas.\cite{2d} One should note that the 2D electron-gas does
not spin-polarize before the transition to a Wigner-crystal, but the
spin-polarized state is very close in energy.\cite{2d} One possible
explanation for the difference could be that the multi-particle
exchange we find in the six-electron dot favors spin-polarization.  In
the 2D electron-gas, such process is less favorable.  To qualitatively
study the transition to a Wigner-molecule-like state, we have
introduced a measure function $\tilde \rho$, which we show to be very
useful for the study of the electron localization.

Overall, we have found VMC to be a perfect tool for studying the
properties of QDs in a wide range of system parameters, resulting
energies in good agreement with the most accurate results available
and enabling us to study the delicate transition of a QD to the
classical regime.

\begin{acknowledgments}
We would like to that M. Alatalo, M. Marlo, H. Saarikoski, and
 V.A. Sverdlov for useful discussions and comments.
\end{acknowledgments}


\begin{thebibliography}{99}
\bibitem{units} We measure the length in units of
$l_0=\sqrt{\frac{\hbar}{m^* \omega}}$.  The scaled strength of the
Coulomb interaction is $C=\sqrt{\frac{\mathrm{ Ha}}{\hbar\omega}}
\frac{\sqrt{m^*/m_0}}{\epsilon}$, where Ha is Hartree.
\bibitem{uns} The numbers in the parentheses are the statistical
uncertainties in the last digit.
\end{thebibliography}
\end{document}